\documentclass{article}

\usepackage{arxiv}
\usepackage{graphicx, color}
\usepackage{amsmath,amssymb}
\usepackage[utf8]{inputenc} 
\usepackage[T1]{fontenc}    
\usepackage{hyperref}       
\usepackage{url}            
\usepackage{booktabs}       
\usepackage{amsfonts}       
\usepackage{microtype}      
\usepackage{lipsum}

\title{Pendulum Analysis by Leaf Functions and Hyperbolic Leaf Functions}

\author{
  Kazunori Shinohara\thanks{10-3 Takiharu-cho, Minami-ku, Nagoya 457-8530, Japan} \\
  Department of Mechanical Systems Engineering\\
  Daido University\\
  10-3 Takiharu-cho, Minami-ku, Nagoya 457-8530, Japan \\
  \texttt{shinohara@06.alumni.u-tokyo.ac.jp} \\
}

\begin{document}
\maketitle

\begin{abstract}  
The mathematical model representing the equation of motion of a pendulum is nonlinear. Solutions that satisfy the equation cannot be represented by elementary functions, such as trigonometric functions. To solve such problems, it is common to linearize the nonlinear equations and derive approximate numerical solutions and exact solutions. Applying such linearization is limited to cases in which the angle of the pendulum is relatively small. In cases where the angle of the pendulum is large, various methods have been presented that rely on numerical solutions and the exact solutions based on Jacobian elliptic functions, to name one example. On the other hand, the author has been studying a certain type differential equation. The second derivative of a function is equal to the term multiplied by $-n$(or $n$) for terms whose function is raised to $2n-1$. Curves based on solving this differential equation are constructed as regular waves with periods. The author has termed the function satisfying this differential equation the leaf function (or the hyperbolic leaf function). In this paper, we attempt to apply this leaf function (or hyperbolic leaf function) to undamped pendulums and overdamped pendulums for large angles.

\end{abstract}


\section{Introduction}
\label{Introduction}
\subsection{Pendulum}
\label{Pendulum}
Not only is the motion of pendulums widely studied in many fields of research, it is also a standard topic utilized to teach dynamics to undergraduate students.
Numerical analysis using various approximate solutions has been proposed to predict the motion.
The basic idea behind the analytical method for nonlinear problems known as PAM(Process Analysis Method) has been previously described and used to obtain the second-order approximate solutions of a pendulum \cite{Liao}. An approximate formula is derived from a logarithmic approximation of the pendulum's period \cite{Lima3}. A closed form approximation of the period of the pendulum has previously been proposed \cite{Lima} \cite{Kim}. A precise formula for the period of a nonlinear pendulum is obtained using the linear delta expansions \cite{Amo}. The arithmetic-geometric mean algorithm has been used to derive a sequence of approximate solutions for the period of a simple pendulum \cite{Car}, and accurate approximate expressions have also been derived for the solution of the equation of motion of a simple pendulum \cite{Bel}. An approximate analytical formula for the period has been obtained using the power series expansion \cite{Bel3} \cite{Jan} \cite{Qin} \cite{Bel4} \cite{Jan2}. An exact solution of the equation of motion of a pendulum using a function has also been proposed. By applying the principle of conservation of energy, exact analytical solutions have been derived using the inverse trigonometric functions and the Jacobi elliptic function\cite{Lima2}. An analytical solution for the differential equation of the nonlinear equation of motion of a pendulum has been derived using the Jacobi elliptic function \cite{Och}. The analytical solution of the damped Helmholtz-Duffing oscillator has also been derived \cite{Alex}.  The exact analytical formulas for the period of a simple pendulum have been obtained in terms of the Jacobi elliptic function \cite{Bel2} \cite{Sal}. 
 One common problem in the aforementioned literature is that, as the initial angle of the pendulum increases, the accuracy of the period decreases. In order to describe the problem in detail, let us consider the problem of the pendulum shown in Fig. 1. The following equation of motion is obtained.

\begin{equation}
\frac{\mathrm{d}^2 \theta}{\mathrm{d}t^2}+\frac{c}{m}\frac{\mathrm{d} \theta}{\mathrm{d}t} +\frac{g}{L}\mathrm{sin}(\theta)=0 \label{1.1.1}.
\end{equation}

The variable $\theta$ represents the pendulum angle, which is a time-dependent function. The parameters $m$, $g$, $L$ and $t$ represent the mass, the gravitational acceleration, the length of the pendulum, and the time, respectively. The Taylor expansion of $\mathrm{sin} (\theta)$ yields the following equation.

\begin{equation}
\mathrm{sin}(\theta)=\theta-\frac{\theta^3}{6}+\frac{\theta^5}{120}- \cdots \label{1.1.2}.
\end{equation}

In this case, since the deflection angle of the pendulum is generally small, we can set $\mathrm{sin}(\theta) \fallingdotseq \theta$. Then, equation (\ref{1.1.1}) becomes the following expression.

\begin{equation}
\frac{\mathrm{d}^2 \theta}{\mathrm{d}t^2}+\frac{c}{m}\frac{\mathrm{d} \theta}{\mathrm{d}t} +\frac{g}{L}\theta=0 \label{1.1.3}.
\end{equation}

\begin{figure*}[tb]
\includegraphics[width=0.75 \textwidth]{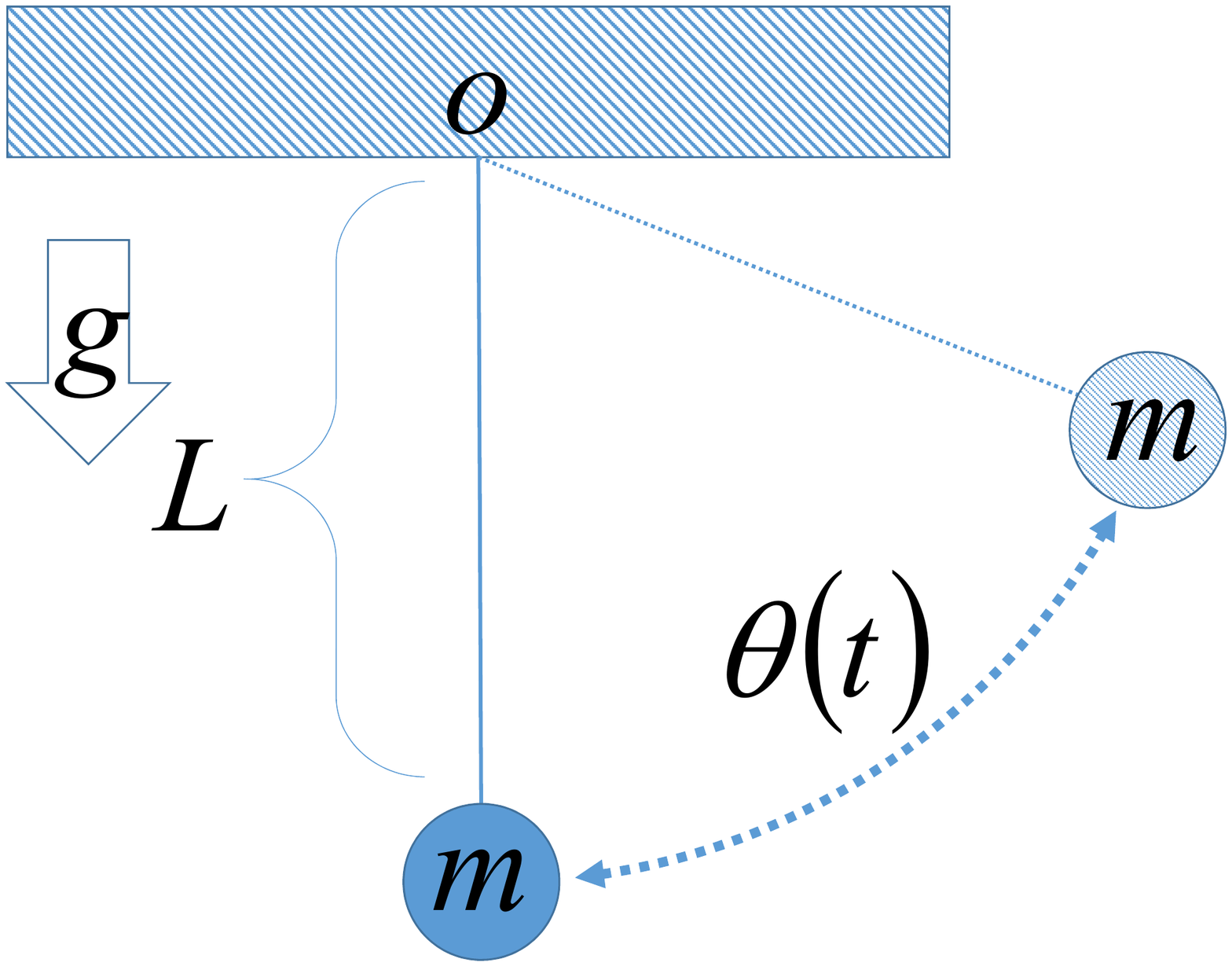}
\caption{Pendulum}
\label{fig:1}
\end{figure*}

\subsection{Purpose}
\label{Purpose}
Let us consider the case where the angle $\theta$ is $\pi/6$.
The values of the first and second terms of the Taylor expansion from Eq. (\ref{1.1.2}) can be obtained as follows.

\begin{equation}
\theta=\frac{\pi}{6}=0.523599 \cdots \label{1.2.1}.
\end{equation}

\begin{equation}
\frac{\theta^3}{6}=\frac{\pi^3}{6^4}=0.0239246 \cdots \label{1.2.2}.
\end{equation}
Compared to the first term $\theta$ of the Taylor expansion  in Eq. (\ref{1.1.2}), the ratio of the second term $\theta^3/6$  is approximately $4.5 \% $. In the engineering field, it is common to assume that the second term can be neglected with respect to the numerical value of the first term if the ratio is within 10\%.
If the deflection angle of the pendulum is small, all terms except the first term in Eq. (\ref{1.1.2}) can be ignored.
 Let us consider the case where the deflection angle is $\pi/2$. The values of the first and second terms of the Taylor expansion in Eq. (\ref{1.1.2}) can be obtained as follows.

\begin{equation}
\theta=\frac{\pi}{2}=1.5708 \cdots \label{1.2.3}.
\end{equation}

\begin{equation}
\frac{\theta^3}{6}=\frac{\pi^3}{6 \cdot 2^3}=0.645964 \cdots \label{1.2.4}.
\end{equation}

Compared to the first term $\theta$ of the Taylor expansion in Eq. (\ref{1.1.2}), the ratio of the second term $\theta^3/6$ is about 40\%. When the deflection angle is large, the second term in Eq. (\ref{1.1.2}) is required. That is, it is necessary to reconsider the following equation of motion.

\begin{equation}
\frac{\mathrm{d}^2 \theta}{\mathrm{d}t^2}+\frac{c}{m}\frac{\mathrm{d} \theta}{\mathrm{d}t} +\frac{g}{L}\left(\theta-\frac{\theta^3}{6} \right)=0 \label{1.2.5}.
\end{equation}

The equation of motion in Eq. (\ref{1.2.5}) is also known as the Duffing equation. Exact solutions of the equation of motion cannot be represented by trigonometric functions.

On the other hand, the author has studied the following ordinary differential equations \cite{Kaz_sl}\cite{Kaz_cl}\cite{Kaz_sh}\cite{Kaz_ch}.

\begin{equation}
\frac{\mathrm{d}^2 \theta}{\mathrm{d}t^2}=-n\theta^{2n-1} \label{1.2.6}.
\end{equation}
\begin{equation}
\theta(0)=1 \label{1.2.7}.
\end{equation}
\begin{equation}
\frac{\mathrm{d}\theta(0)}{\mathrm{d}t}=0 \label{1.2.8}.
\end{equation}

The variable $\theta$ in the above equations represents an unknown function. The unknown function depends on the time $t$. The term $\mathrm{d}\theta/\mathrm{d}t$ and the term $\mathrm{d}^2\theta/\mathrm{d}t^2$ represent the first and second derivatives of $\theta$, respectively. Eq. (\ref{1.2.6}) is a very simple differential equation. Despite being extremely simple, as can be seen in Eq. (\ref{1.2.6}), solving ordinary differential equations for numerical analysis yields waves for arbitrary $n$. Therefore, trigonometric functions can generally be considered for exact solutions that satisfy ordinary differential equations. However, the exact solution of the above equation cannot be obtained using trigonometric functions. The author believed that the fundamental function required to derive the exact solution of the ordinary differential equation satisfies the above differential equation. This function was defined as follows.

\begin{equation}
\theta(t)=\mathrm{cleaf}_{n}(t) \label{1.2.9}.
\end{equation}

This function is called the leaf function. The symbol ``cleaf'' represents the combination of both ``cos'' and ``leaf.'' Let us consider another differential equation.

\begin{equation}
\frac{\mathrm{d}^2\theta}{\mathrm{d}t^2}=n\theta^{2n-1} \label{1.2.10}.
\end{equation}
\begin{equation}
\theta(0)=0 \label{1.2.11}.
\end{equation}
\begin{equation}
\frac{\mathrm{d}\theta(0)}{\mathrm{d}t}=1 \label{1.2.12}.
\end{equation}

A function satisfying the above differential equation is defined as follows.

\begin{equation}
\theta(t)=\mathrm{\mathrm{sleafh}}_{n}(t) \label{1.2.13}.
\end{equation}

This function is defined as the hyperbolic leaf function. In this paper, we aim to obtain solutions with high accuracy even if the deflection angle of the pendulum is large. The exact solutions that satisfy the Duffing equation in Eq. (\ref{1.2.5}) are represented by the leaf function or the hyperbolic leaf function. Two types of exact solutions are applied to the motion equation of the pendulum, one type to the undamped pendulum and the other to the overdamped pendulum. Exact solutions based on the leaf function or the hyperbolic leaf function are verified by comparing them with both the non-approximate equation from Eq. (\ref{1.1.1}) and  the trigonometric function solution from Eq. (\ref{1.1.3}).

\section{Exact solutions using leaf function and hyperbolic leaf function}
\subsection{Exact solution of undamped pendulum}
The exact solutions of the leaf function used in this paper are now described. 
In case where the damping coefficient $c=0$, the equation of motion of the undamped pendulum is as follows.

\begin{equation}
\frac{\mathrm{d}^2 \theta}{\mathrm{d}t^2} +\frac{g}{L}\left(\theta-\frac{\theta^3}{6} \right)=0 \label{2.1.1}.
\end{equation}

The exact solution satisfying this equation of motion is the following equation\cite{Kaz_duf}.

\begin{equation}
\theta(t)=A \mathrm{sin} \left(B \int_{0}^{\omega t + \phi} \mathrm{cleaf}_2(Bu)\mathrm{d}u  \right) \label{2.1.2}.
\end{equation}
The first derivative of the above equation is obtained as follows.

\begin{equation}
\frac{\mathrm{d}\theta}{\mathrm{d}t}=A\mathrm{cos} \left(B \int_{0}^{\omega t+\phi} \mathrm{cleaf}_2(B u)\mathrm{d}u  \right)  \cdot B \cdot \omega \cdot \mathrm{cleaf}_2(B \omega t+ B \phi) \label{2.1.3}.
\end{equation}

The second derivative of the above equation is obtained as follows.

\begin{equation}
\begin{split}
\frac{\mathrm{d}^2 \theta}{\mathrm{d}t^2}=-A \mathrm{sin} \left(B \int_{0}^{\omega t+\phi} \mathrm{cleaf}_2(B u)\mathrm{d}u  \right) \cdot B^2 \cdot \omega^2 \cdot (\mathrm{cleaf}_2(B \omega t+ B \phi))^2 \\
 - A \mathrm{cos} \left(B \int_{0}^{\omega t+\phi} \mathrm{cleaf}_2(B u)\mathrm{d}u \right) \cdot B^2 \cdot \omega^2 \cdot \sqrt{1-(\mathrm{cleaf}_2(B \omega t + B \phi))^4 }
\label{2.1.4}.
\end{split}
\end{equation}

Substituting Eq. (\ref{A4}) into the above equation yields the following equation.

\begin{align}
\begin{split}
\frac{d ^2 \theta}{\mathrm{d}t^2}  &=  - A \mathrm{sin} \left( B \int_{0}^{\omega t+\phi} \mathrm{cleaf}_2(B u)\mathrm{d}u   \right) \cdot B^2 \cdot \omega^2 \cdot \mathrm{cos} \left( 2B \int_{0}^{\omega t+\phi} \mathrm{cleaf}_2(Bu)\mathrm{d}u \right) \\
&- A \mathrm{cos} \left(B \int_{0}^{\omega t+\phi} \mathrm{cleaf}_2(Bu)\mathrm{d}u  \right) 
\cdot B^2 \cdot \omega^2 \cdot \mathrm{sin} \left(2 B \int_{0}^{\omega t+\phi} \mathrm{cleaf}_2(Bu)\mathrm{d}u \right) \\
& = -A B^2 \omega^2 \mathrm{sin} \left(3B \int_{0}^{\omega t+\phi} \mathrm{cleaf}_2(u)\mathrm{d}u \right) \\
& = - 3 B^2 \omega^2 A \mathrm{sin} \left(B \int_{0}^{\omega t+\phi} \mathrm{cleaf}_2(Bu)\mathrm{d}u  \right) \\
& + 4 \frac{B^2 \omega^2}{A^2} 
\left\{ A \mathrm{sin} \left( B \int_{0}^{\omega t+\phi} \mathrm{cleaf}_2(Bu)\mathrm{d}u   \right) \right\}^3 
   \label{2.1.5}.
\end{split}
\end{align}

Substituting Eq. (\ref{2.1.2}) into the above equation leads to the following equation.

\begin{align}
\begin{split}
\frac{\mathrm{d}^2 \theta}{\mathrm{d}t^2}  + 3 B^2 \omega^2 \theta - 4 \frac{B^2 \omega^2}{A^2} \theta^3=0 \label{2.1.6}.
\end{split}
\end{align}

The initial conditions can be obtained using the following equations.

\begin{equation}
 \theta(0)=A \mathrm{sin}\left(B \int_{0}^{\phi} \mathrm{cleaf}_2(B u)\mathrm{d}u \right) 
\label{2.1.7}.
\end{equation}
\begin{equation}
\frac{\mathrm{d} \theta(0)}{\mathrm{d}t}=AB \omega \mathrm{cos}\left(B \int_{0}^{\phi} \mathrm{cleaf}_2(Bu)\mathrm{d}u \right) \cdot \mathrm{cleaf}_2(B \phi)
\label{2.1.8}.
\end{equation}

\subsection{Exact solution of overdamped pendulum}
The overdamped pendulum is strongly influenced by damping, such that the mass of the pendulum does not vibrate. It moves only from the initial position to the final position. The exact solution of the overdamped pendulum satisfying Eq. (\ref{1.2.5}) is given by the following equation.

\begin{align}
\begin{split}
\theta(t)=A \mathrm{e}^{D \omega t} \mathrm{\mathrm{sleafh}}_2(B \mathrm{e}^{D \omega t}+\phi) \label{2.2.1}.
\end{split}
\end{align}

The first derivative of the above equation is obtained as follows.

\begin{align}
\begin{split}
\frac{\mathrm{d}\theta}{\mathrm{d}t}
&=AD \omega \mathrm{e}^{D\omega t} \mathrm{\mathrm{sleafh}}_2(B \mathrm{e}^{D \omega t} +\phi) 
\\
&+A B D \omega \mathrm{e}^{2 D \omega t} 
\cdot \sqrt{1+(\mathrm{\mathrm{sleafh}}_2(B\mathrm{e}^{D \omega t}+\phi))^4}
\label{2.2.2}.
\end{split}
\end{align}

The second derivative of the above equation is obtained as follows.

\begin{align}
\begin{split}
\frac{\mathrm{d}^2\theta}{\mathrm{d}t^2}
&=A D^2 \omega^2 \mathrm{e}^{D \omega t} \mathrm{sleafh}_2(B \cdot \mathrm{e}^{D \omega t} +\phi) \\
& +A D \omega \mathrm{e}^{D \omega t} \sqrt{1+(\mathrm{sleafh}_2(B\mathrm{e}^{D \omega t}+\phi))^4}
(B D \omega \mathrm{e}^{D \omega t} ) \\
&+ 2 A B D^2 \omega^2 \mathrm{e}^{2 D \omega t} \sqrt{1+(\mathrm{sleafh}_2(B\mathrm{e}^{D \omega t}+\phi))^4} \\
&+2 A B D \omega \mathrm{e}^{2 D \omega t} (\mathrm{sleafh}_2(B\mathrm{e}^{D \omega t}+\phi))^3 \cdot (B D \omega \cdot \mathrm{e}^{D \omega t}) \\
&=A D^2 \omega^2 \mathrm{e}^{D \omega t} \mathrm{sleafh}_2(B \cdot \mathrm{e}^{D \omega t} +\phi) \\
&+2A B^2 D^2 \omega^2 \mathrm{e}^{3 D \omega t} (\mathrm{sleafh}_2(B \mathrm{e}^{D \omega t} +\phi))^3 \\
&+3AB D^2 \omega^2 \mathrm{e}^{2 D \omega t} \sqrt{1+(\mathrm{sleafh}_2(B\mathrm{e}^{D \omega t}+\phi))^4} \label{2.2.3}.
\end{split}
\end{align}

Substituting Eqs. (\ref{2.2.1}) and (\ref{2.2.2}) into Eq. (\ref{2.2.3}) yields the following equation.

\begin{equation}
\frac{\mathrm{d}^2 \theta}{\mathrm{d}t^2} -3 D \omega \frac{\mathrm{d} \theta}{\mathrm{d}t}+ 2 D^2 \omega^2 \theta -2 \left(\frac{B D \omega}{A}\right)^2 \theta^3=0
\label{2.2.4}.
\end{equation}

The initial conditions for the above equation are as follows.
\begin{equation}
\theta(0)=A \mathrm{sleafh}_2(B+\phi) \label{2.2.5}.
\end{equation}
\begin{equation}
\frac{\mathrm{d} \theta(0)}{\mathrm{d}t}=A D \omega \mathrm{sleafh}_2(B+\phi)
+A B D \omega \sqrt{1+(\mathrm{sleafh}_2(B+\phi))^4} \label{2.2.6}.
\end{equation}

\section{Analysis}
\subsection{Initial conditions}
In this paper, we verify the accuracy in the case of approximating $\mathrm{sin}(\theta) \fallingdotseq \theta$,  $\mathrm{sin}(\theta) \fallingdotseq \theta - \theta^ 3/6$, as well as in the case of not approximating $\mathrm{sin} (\theta)$ in the ordinary differential equation (\ref{1.1.1}). Therefore, the initial conditions are set such that the coefficients of the differential equation based on the Duffing equation of motion agree with those obtained from the leaf function or the hyperbolic leaf function. The parameters of $\omega$ are set as follows.

\begin{equation}
\omega = \sqrt{\frac{g}{L}}  \label{3.1.3}.
\end{equation}

The initial conditions of the mass are set as follows.

\begin{equation}
\theta(0) = 2  \label{3.1.4}.
\end{equation}

\begin{equation}
\frac{\mathrm{d}\theta(0)}{\mathrm{d}t}=0  \label{3.1.5}.
\end{equation}

The mass is moved from the final position to the initial position  $\theta(0) = 2/\pi \times 180 \fallingdotseq 114.6^\circ$ . To simplify the analysis, the length and mass of the pendulum are set to $L = 10.0 $ and $m=1.0$, respectively. The parameter $g$ represents the gravitational acceleration, which takes on a value of $10.0$.

\subsection{Undamped pendulum}
The results of numerical analysis of the undamped pendulum are discussed in this section. The parameters A and B are set as follows, so as to match the coefficients in Eq. (\ref{2.1.6}) with those in Eq. (\ref{2.1.1}).

\begin{equation}
A= 2 \sqrt{2}  \label{3.1.1}.
\end{equation}

\begin{equation}
B = \frac{1}{\sqrt{3}}  \label{3.1.2}.
\end{equation}

When set as above, the coefficients $\theta$ and $\theta^3$ from the Taylor expansion $\mathrm{sin}(\theta)$ perfectly match those in Eq. (\ref{2.1.1}). The differential equations and exact solutions represent non-approximate solutions of $\mathrm{sin} (\theta)$, the first-order approximation of $\mathrm{sin} (\theta)$ and the second-order approximation of $\mathrm{sin} (\theta)$, respectively.

$\cdot$ Non-approximate (ODE)
\begin{equation}
\frac{\mathrm{d}^2 \theta}{\mathrm{d}t^2}+\mathrm{sin}(\theta)=0 \label{3.2.1}.
\end{equation}

$\cdot$ First-order approximation of $\mathrm{sin} (\theta$) (ODE and exact solution)
\begin{equation}
\frac{\mathrm{d}^2 \theta}{\mathrm{d}t^2}+\theta=0 \label{3.2.2}.
\end{equation}

\begin{equation}
\theta(t)=2 \mathrm{sin} \left( t+\frac{\pi}{2} \right) \label{3.2.3}.
\end{equation}

$\cdot$ Second-order approximation of $\mathrm{sin} (\theta)$ (ODE and exact solution)

\begin{equation}
\frac{\mathrm{d}^2 \theta}{\mathrm{d}t^2} + \theta-\frac{\theta^3}{6} =0 \label{3.2.4}.
\end{equation}

\begin{equation}
\theta(t)=2 \sqrt{2} 
\mathrm{sin} \left(\frac{1}{\sqrt{3}} \int_{0}^{ t + \frac{\sqrt{3}}{2} \pi_2 }  
\mathrm{cleaf}_2 \left( \frac{1}{\sqrt{3}} u \right)\mathrm{d}u \right) \label{3.2.5}.
\end{equation}

For more details, refer to Appendix B. The curves obtained from these equations are shown in Fig. 2. The horizontal axis represents time $t$ and the vertical axis represents the angle $\theta$. The initial conditions of these curves are set as in Eq. (\ref{3.1.4}) and Eq. (\ref{3.1.5}). The curve obtained from Eq. (\ref{3.2.1}) is compared to the curve obtained from Eq. (\ref{3.2.5}).  As shown in the graph, the curves obtained from Eq. (\ref{3.2.1}) and Eq. (\ref{3.2.5}) almost agree in the region of the first period. In Eq. (\ref{3.2.3}), the error accumulates as time progress. The wave given by Eq. (\ref{3.2.3}) gradually shifts from the wave given by Eq.  (\ref{3.2.1}). The periods obtained from Eq. (\ref{3.2.1}), Eq. (\ref{3.2.3}) and Eq. (\ref{3.2.5}) are shown in Table 1. The rounding errors (\%) for one cycle resulting from Eq. (\ref{3.2.1}) are also shown in Table 1.
Although the error of one period obtained from Eq. (\ref{3.2.2}) is 24.8\%, the error of one period obtained from Eq. (\ref{3.2.5}) is 8.8\%. This shows that the accuracy of the solution is improved when the leaf function is used.

\begin{figure*}[tb]
\includegraphics[width=0.75 \textwidth]{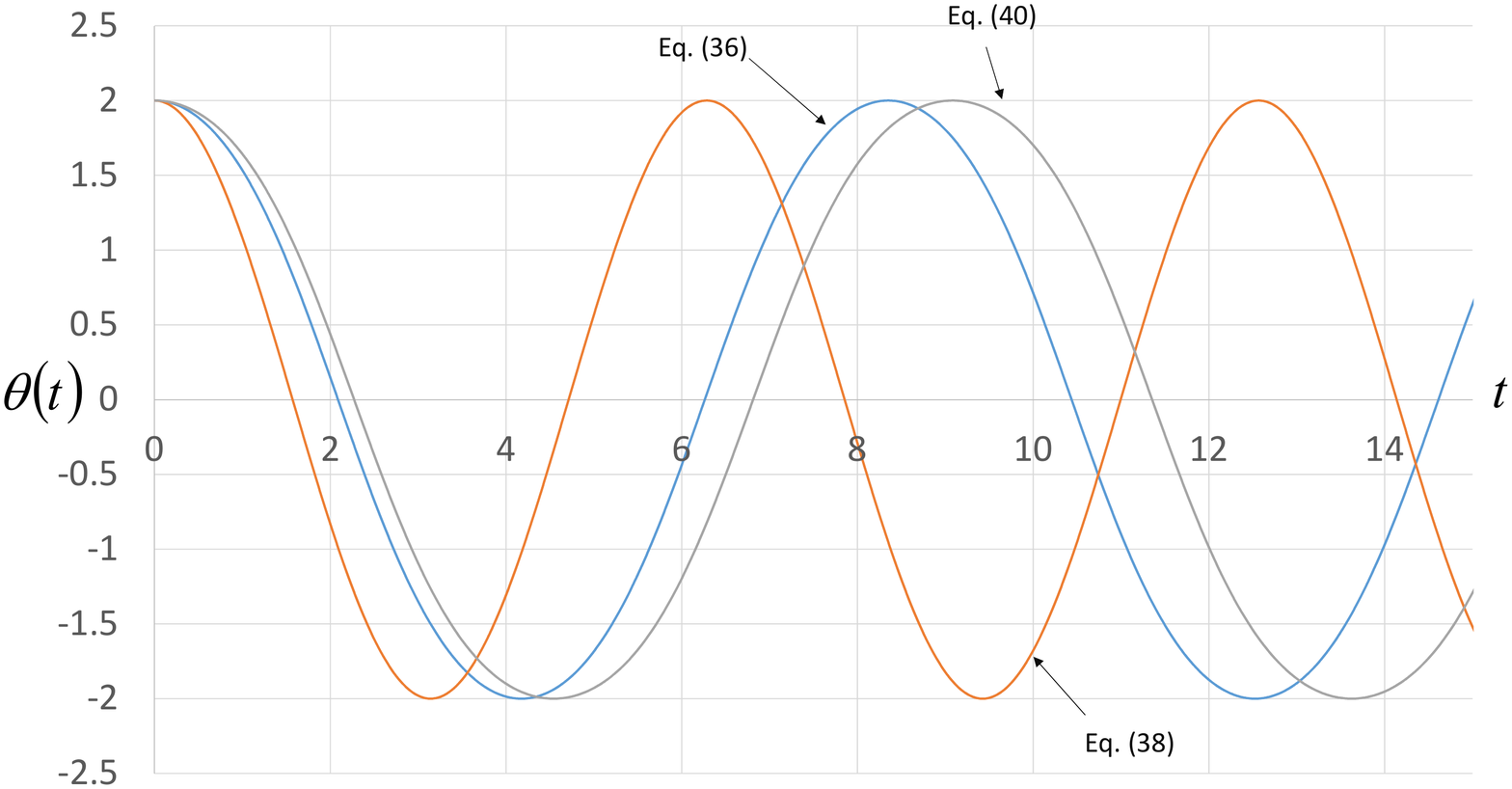}
\caption{Curves obtained from Eq. (\ref{3.2.1}), (\ref{3.2.3}) and (\ref{3.2.5}) (the horizontal and vertical axes represent time $t$ and angle $\theta$, respectively)}
\label{fig:2}
\end{figure*}

\begin{table}
\caption{ One period of wave }
\label{tab:1} 
\begin{tabular}{ccc}
\hline\noalign{\smallskip}
 -  & Period & 
\begin{tabular}{c}
Error (\%) with respect to \\ period obtained from Eq. (\ref{D7})  \end{tabular}\\ 
\noalign{\smallskip}\hline\noalign{\smallskip}
Numerical value obtained from Eq. (\ref{D7})	&8.349 $\cdots$	&	-	\\
Numerical value obtained from Eq. (\ref{D5})	&	6.283 $\cdots$	&	24.8	\\
Numerical value obtained from Eq. (\ref{D6})	&	9.083 $\cdots$	&	8.8	\\
\noalign{\smallskip}\hline
\end{tabular}
\end{table}

\subsection{Overdamped pendulum}
The results of numerical analysis of the overdamped pendulum will now discussed.
An exact solution of the Duffing equation of motion can be described using the hyperbolic leaf function $\mathrm{sleafh}_n(t)$ \cite{Kaz_sh}.
The symbol $\mathrm{sleafh}_n(t)$ is an artificially defined function analogous to $\mathrm{sinh}(t)$. Waves are not generated by this function. $\mathrm{sleafh}_n(t)$ is a function that monotonically increases or decreases as time progress. The damping coefficient is set such that the curve of the hyperbolic leaf function can be computed using a second-order approximation, which can then be compared to the curve obtained without approximation and the curve obtained from the first-order approximation of the function sin$(\theta)$. The coefficients of the Taylor expansion $\theta-\theta^3/6$ in Eq. (\ref{1.2.5}) perfectly match those from Eq. (\ref{2.2.4}) under the initial conditions given by Eqs. (\ref{3.1.4}) and (\ref{3.1.5}). The differential equations and exact solutions based on the non-approximate Taylor expansion of the first-order approximation of sin($\theta$) and the second-order approximation of sin($\theta$) are as follows:

$\cdot$ Non-approximate (ODE)
\begin{equation}
\frac{\mathrm{d}^2 \theta}{\mathrm{d}t^2}+\frac{3}{\sqrt{2}}  \frac{\mathrm{d} \theta}{\mathrm{d}t}+\mathrm{sin}(\theta)=0 \label{3.3.1}.
\end{equation}

$\cdot$ First-order approximation of $\mathrm{sin} (\theta)$ (ODE and exact solution)
\begin{equation}
\frac{\mathrm{d}^2 \theta}{\mathrm{d}t^2}+\frac{3}{\sqrt{2}}  \frac{\mathrm{d} \theta}{\mathrm{d}t}+ \theta=0 \label{3.3.2}.
\end{equation}

\begin{equation}
\theta(t)=-2 \mathrm{e}^{-\sqrt{2}t}+4 \mathrm{e}^{-\frac{1}{\sqrt{2}}t} \label{3.3.3}.
\end{equation}

$\cdot$ Second-order approximation of $\mathrm{sin} (\theta)$ (ODE and exact solution)

\begin{equation}
\frac{\mathrm{d}^2 \theta}{\mathrm{d}t^2} +\frac{3}{\sqrt{2}}  \frac{\mathrm{d} \theta}{\mathrm{d}t}+   \theta-\frac{\theta^3}{6} =0 \label{3.3.4}.
\end{equation}

\begin{equation}
\theta(t)=2^{\frac{3}{4}} \mathrm{e}^{-  \frac{1}{\sqrt{2}} t   } \cdot
\mathrm{sleafh}_2 \left( -\frac{\sqrt[4]{2}}{\sqrt{3}} \cdot 
\mathrm{e}^{- \frac{1}{\sqrt{2}} t    }  + \phi \right) \label{3.3.5}.
\end{equation}

Refer to the Appendix C for additional details. The curves corresponding to the above equations are shown in Fig. 3.
The horizontal and vertical axes represent time and the angle, respectively. Since the motion is significantly affected by damping, the angle $\theta$ gradually decreases from its initial position in Eq. (\ref{3.1.4}). The mass ultimately moves to the final position, after which it does not vibrate. As shown in the graph, the curve of the non-approximate solution obtained from Eq. (\ref{3.3.1}) nearly agrees with the hyperbolic leaf curve obtained from Eq. (\ref{3.3.5}). On the other hand, the equation of motion obtained using the first-order Taylor expansion demonstrates differences in the transition from the initial position (\ref{3.1.4}) to the final position $\theta=0$. At approximately 10.0 seconds, which corresponds to the time at which the pendulum will have reached the final position, the three curves are nearly in agreement with $\theta=0$.

\begin{figure*}[tb]
\includegraphics[width=0.75 \textwidth]{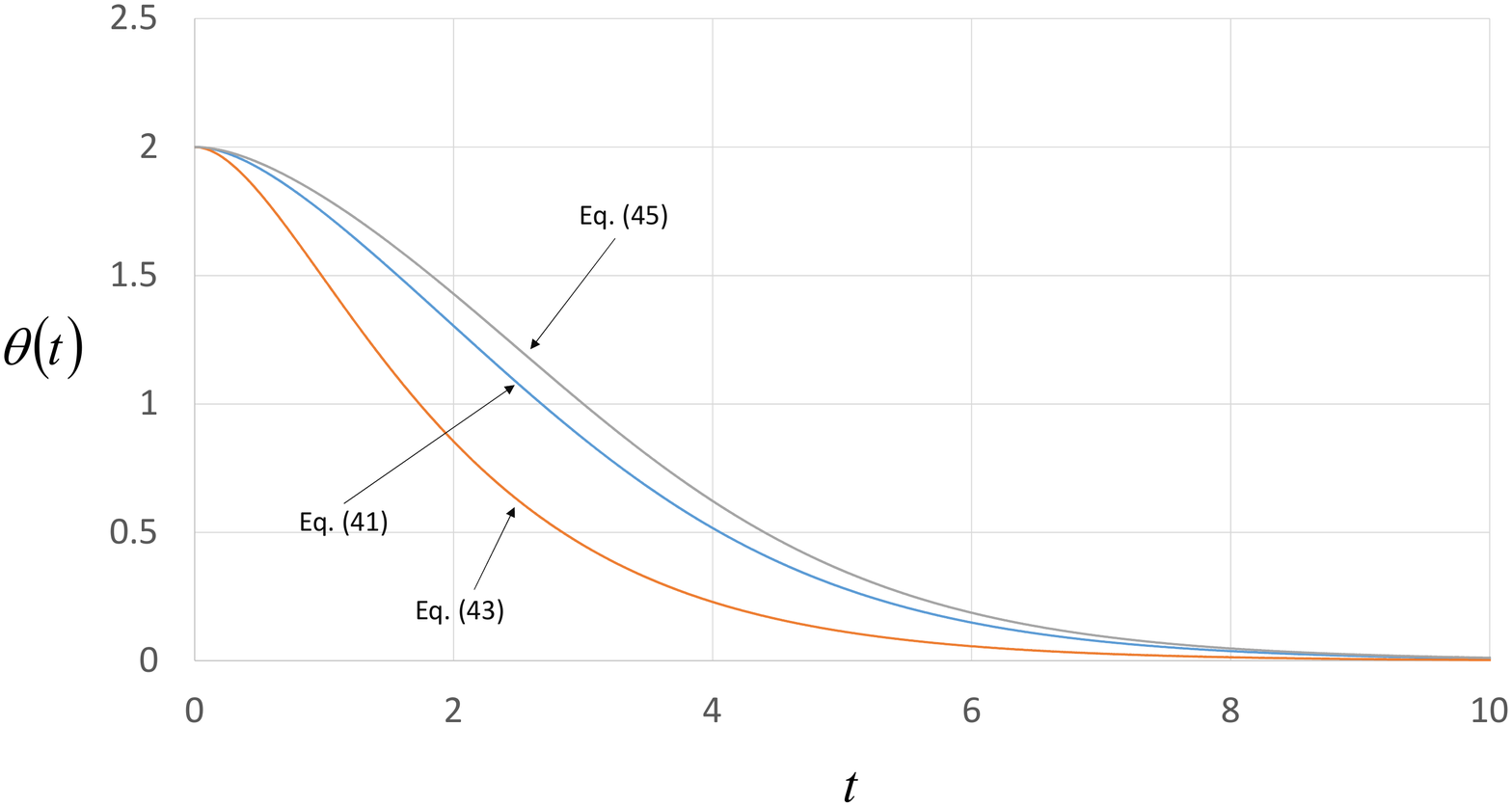}
\caption{Curves obtained from Eq. (\ref{3.3.1}), (\ref{3.3.3}) and (\ref{3.3.5}) (the horizontal and vertical axes represent time $t$ and angle $\theta$, respectively)}
\label{fig:3}
\end{figure*}

\subsection{Damped pendulum}
Using the hyperbolic leaf function $\mathrm{sleafh}_n(t)$, the damping coefficient of the term $\mathrm{d} \theta/\mathrm{d}t$ cannot be manipulated by the parameters of the function.
The leaf functions make it possible to derive a differential equation consisting of the term $\mathrm{d}^2 \theta/\mathrm{d}t^2$,  the term $\mathrm{d} \theta/\mathrm{d}t$, the term $\theta$, and the term $\theta^3/6$, as shown in Eq. (\ref{1.2.5}). However, comparing the ODE obtained from the leaf function in Eq. (\ref{1.2.5}), it is observed that the coefficients of each term in the ODE cannot be perfectly matched. Hence, they are not described in this paper.
The curve obtained from Eq. (\ref{3.2.5}) represents the damping coefficient $c=0$ in Eq. (\ref{1.2.5}). The curve obtained from Eq. (\ref{3.3.5}) represents the curve of the overdamped pendulum. Therefore, if the influence of the damping term $\mathrm{d} \theta/\mathrm{d}t$ is reduced, the wave is damped during vibration. The coefficients of the damped term in Eqs. (\ref{3.3.1}), (\ref{3.3.2}) and (\ref{3.3.4}) are each set to 2/3. Fig. 4 shows the analysis of the graph.

$\cdot$ Non-approximate (ODE)
\begin{equation}
\frac{\mathrm{d}^2 \theta}{\mathrm{d}t^2}+\frac{2}{3} \frac{3}{\sqrt{2}} \frac{\mathrm{d} \theta}{\mathrm{d}t}+\mathrm{sin}(\theta)=0 \label{3.4.1}.
\end{equation}

$\cdot$ First-order approximation of $\mathrm{sin} (\theta)$ (ODE and exact solution)
\begin{equation}
\frac{\mathrm{d}^2 \theta}{\mathrm{d}t^2}+\frac{2}{3} \frac{3}{\sqrt{2}}  \frac{\mathrm{d} \theta}{\mathrm{d}t}+ \theta=0 \label{3.4.2}.
\end{equation}

\begin{equation}
\theta(t)=2 \sqrt{2} \mathrm{e}^{-\frac{1}{\sqrt{2}}  t } 
\mathrm{sin} \left( \frac{t}{\sqrt{2}} +  \frac{\pi}{4} \right) \label{3.4.3}.
\end{equation}

$\cdot$ Second-order approximation of $\mathrm{sin} (\theta)$ (ODE and exact solution)

\begin{equation}
\frac{\mathrm{d}^2 \theta}{\mathrm{d}t^2} +\frac{2}{3} \frac{3}{\sqrt{2}}
 \frac{\mathrm{d} \theta}{\mathrm{d}t}+  \theta-\frac{\theta^3}{6} =0 \label{3.4.4}.
\end{equation}

The coefficients of the differential equations in Eqs. (\ref{3.4.1}), (\ref{3.4.2}) and (\ref{3.4.4}) are identical to the coefficients of the differential equations in Eqs. (\ref{3.3.1}), (\ref{3.3.2}) and (\ref{3.3.4}), with the exception of the damping term $\mathrm{d} \theta/\mathrm{d}t $. As the damping effect decreases, the pendulum begins to vibrate. If the term $\mathrm{d} \theta/\mathrm{d}t$ is completely eliminated, Eqs. (\ref{3.4.1}), (\ref{3.4.2}) and (\ref{3.4.4}) completely match Eqs. (\ref{3.2.1}), (\ref{3.2.2}) and (\ref{3.2.4}), respectively. The motion of the pendulum shifts from that of the damped system to that of the undamped system.

\begin{figure*}[tb]
\includegraphics[width=0.75 \textwidth]{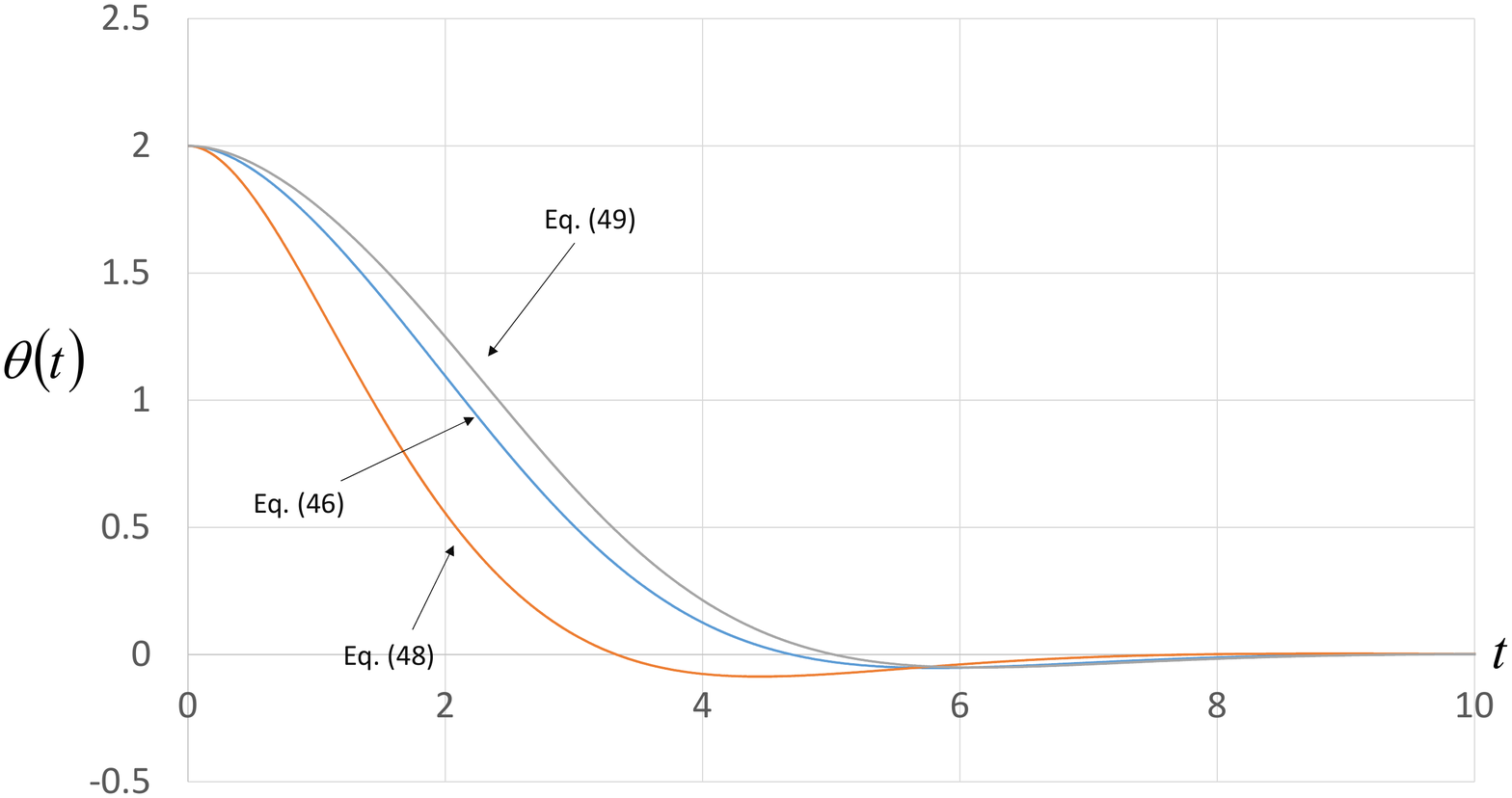}
\caption{Time history of damped pendulum (the horizontal and vertical axes represent time $t$ and angle $\theta$, respectively)}
\label{fig:4}
\end{figure*}

\section{Conclusion}
\label{Conclusion}
In this paper, the leaf function and the hyperbolic leaf function are applied to the study of the motion of the pendulum. To derive the exact solution of the equation of motion of the pendulum, the equation must approximate $\mathrm{sin} (\theta) \fallingdotseq \theta$. This makes it possible to derive the exact solution of the equation of motion using trigonometric functions. An approximation of $\mathrm{sin} (\theta) \fallingdotseq \theta$ can be obtained under the assumption that the angle is small. The Duffing equation of motion of the pendulum is a differential equation derived by including the second term of the Taylor expansion of $\mathrm{sin} (\theta) ( \fallingdotseq \theta - \theta^3/6 )$. Since the differential equation contains the nonlinear term $\theta^3$, an exact solution cannot be derived with trigonometric functions.
In this study, we attempted to describe the exact solutions by using the leaf function and the hyperbolic leaf function. The following conclusions can be drawn.

$\cdot$ The leaf functions are applied to the Duffing equation of motion of a pendulum. Exact solutions of the undamped Duffing equations and the overdamped Duffing equations are proposed using leaf functions.

$\cdot$ The exact solution of the overdamped Duffing equation of motion monotonically decreases with time.
The exact solution of the ordinary differential equation can be constructed using the hyperbolic leaf function. As the damping parameter in the equation decreases, damped waves are produced. In the case where the damping term is zero, the ordinary differential equation is the same as the undamped Duffing equation of motion of the pendulum. The exact solution of the ordinary differential equation can be constructed using the leaf function. The author found the exact solutions representing the damped wave between the undamped and overdamped Duffing equations of motion. However, the coefficients of the ordinary differential equation based on the exact solution do not perfectly match the coefficients of the damped Duffing equation. Therefore, only waves were described by the numerical analysis.

$\cdot$ The different periods obtained from the undamped equation of motion of the pendulum are compared. The initial angle was set to 114.6 degrees. Compared to the period of the solution without approximation, the error due to the trigonometric function and the error due to the leaf function are $24.8\%$ and $8.8\%$, respectively. The leaf function makes it possible to improve the accuracy even when a large initial angle is chosen.

\appendix
\def\thesection{Appendix}
\section{A}
\label{Appendix A}
The equation relating the leaf functions and the trigonometric functions is described in this section. The following inverse trigonometric function is first differentiated.

\begin{equation}
\begin{split}
& \frac{d}{\mathrm{d}t} \mathrm{arccos}(\mathrm{cleaf}_2(Bt))^2 \\
&=-\frac{1}{\sqrt{1-(\mathrm{cleaf}_2(Bt))^4}} \cdot B \cdot 2 \cdot \mathrm{cleaf}_2(Bt) \left\{-\sqrt{1-(\mathrm{cleaf}_2(Bt))^4} \right\} \\
&=2B \cdot \mathrm{cleaf}_2(Bt)
\label{A1}.
\end{split}
\end{equation}

By integrating the above equation from $0$ to $\omega t+\phi$, the following equation is obtained.

\begin{equation}
\begin{split}
\left[ \mathrm{arccos}(\mathrm{cleaf}_2(Bu))^2 \right]^{\omega t+\phi}_{0}
=\int_0^{\omega t+\phi} 2B \cdot \mathrm{cleaf}_2(Bu) \mathrm{d}u 
=2B \int_0^{\omega t+\phi} \mathrm{cleaf}_2(Bu) \mathrm{d}u 
\label{A2}.
\end{split}
\end{equation}

The left side can be expanded as follows.

\begin{equation}
\begin{split}
& \left[ \mathrm{arccos}(\mathrm{cleaf}_2(Bu))^2 \right]^{\omega t+\phi}_{0}
=\mathrm{arccos}(\mathrm{cleaf}_2(B\omega t+B\phi))^2-\mathrm{arccos}(\mathrm{cleaf}_2(0))^2 \\
&=\mathrm{arccos}(\mathrm{cleaf}_2(B\omega t+B\phi))^2-\mathrm{arccos}(1)
=\mathrm{arccos}(\mathrm{cleaf}_2(B\omega t+B\phi))^2
\label{A3}.
\end{split}
\end{equation}

Using Eqs. (\ref{A2}) and (\ref{A3}), the following relation is obtained.

\begin{equation}
\begin{split}
(\mathrm{cleaf}_2(B\omega t+B\phi))^2
=\mathrm{cos} \left( 2B \int_0^{\omega t+\phi} \mathrm{cleaf}_2(Bu) \mathrm{d}u  \right)
\label{A4}.
\end{split}
\end{equation}

\appendix
\def\thesection{Appendix}
\section{B}
\label{B}

Eq. (\ref{3.2.5}) is now discussed. The parameters are determined by Eqs. (\ref{3.1.1})-(\ref{3.1.2}), as well as the initial conditions (\ref{3.1.4}) and (\ref{3.1.5}). Substituting these parameters into Eq. (\ref{2.1.8}) yields the following equation.

\begin{equation}
\frac{\mathrm{d} \theta(0)}{\mathrm{d}t}=2 \sqrt{2} \frac{1}{\sqrt{3}} 
\mathrm{cos} \left( \frac{1}{\sqrt{3}} \int_{0}^{\phi} cleaf_2 
\left( \frac{1}{\sqrt{3}}u \right)\mathrm{d}u \right) 
\cdot \mathrm{cleaf}_2 \left( \frac{1}{\sqrt{3}} \phi \right) 
\label{B1}.
\end{equation}

Using the equation $\mathrm{cleaf}_2(\pi_2/2)=0$, the parameter $\phi$ is obtained from the following equation.

\begin{equation}
\frac{1}{\sqrt{3}} \phi=\frac{\pi_2}{2} 
\label{B2}.
\end{equation}

Refer to Appendix E for the constant $\pi_2$. Next, the following relation is obtained from Eq. (\ref{2.1.7}).

\begin{equation}
\begin{split}
&\theta(0)=2 \sqrt{2} 
\mathrm{sin} \left( \frac{1}{\sqrt{3}} \int_{0}^{ \frac{\sqrt{3}}{2} \pi_2 } \mathrm{cleaf}_2 
\left( \frac{1}{\sqrt{3}}u \right)\mathrm{d}u  \right) \\
&=2 \sqrt{2} 
\mathrm{sin} \left( \frac{1}{\sqrt{3}} \cdot  \frac{\sqrt{3} \pi}{4}  \right)
=2
\label{B3}.
\end{split}
\end{equation}

Refer to Appendix F for the integral of the above equation.

\appendix
\def\thesection{Appendix}
\section{C}
\label{C}
We now discuss the relationship between the coefficients in the hyperbolic leaf function and  the coefficients in the ODE used in section 2.2.  Substituting $D=-1/\sqrt{2}$ into Eq. (\ref{2.2.4}) yields the following equation.

\begin{equation}
\frac{\mathrm{d}^2 \theta}{\mathrm{d}t^2}+\frac{3}{\sqrt{2}} \omega \frac{\mathrm{d} \theta}{\mathrm{d}t} +\omega^2 \left(\theta-\frac{B^2}{A^2} \theta^3 \right)=0 \label{C1}.
\end{equation}

Comparing the coefficients of each term in Eq. (\ref{1.2.5}) with the coefficients of each term in Eq. (\ref{C1}), the following equation is obtained.

\begin{equation}
c=\frac{3}{\sqrt{2}} m \omega =\frac{3}{\sqrt{2}} m \sqrt{ \frac{g}{L} } 
\label{C2}.
\end{equation}

\begin{equation}
\frac{B^2}{A^2}=\frac{1}{6}  \label{C3}.
\end{equation}

Using the initial condition given by Eq. (\ref{3.1.4}), the following equation is obtained from Eq. (\ref{2.2.5}).

\begin{equation}
\theta(0)=A \cdot \mathrm{sleafh}_2(B+\phi)=-\sqrt{6} B \cdot \mathrm{sleafh}_2(B+\phi)=2 \label{C4}.
\end{equation}

\begin{equation}
\mathrm{sleafh}_2(B+\phi)=-\frac{2}{ \sqrt{6} B} \label{C5}.
\end{equation}

The following equation is obtained from the initial condition given by Eq. (\ref{3.1.5}). 

\begin{equation}
\frac{\mathrm{d} \theta(0)}{\mathrm{d}t}=A D \omega \mathrm{sleafh}_2(B+\phi)
+A B D \omega \sqrt{1+(\mathrm{sleafh}_2(B+\phi))^4}=0 \label{C6}.
\end{equation}

By substituting (\ref{C5}) into (\ref{C6}), the following equation is obtained.

\begin{equation}
\mathrm{sleafh}_2(B+\phi) + B \sqrt{1+(\mathrm{sleafh}_2(B+\phi))^4}=0 \label{C7}.
\end{equation}

\begin{equation}
-\frac{2}{ \sqrt{6} B} + B \sqrt{1+ \left( -\frac{2}{\sqrt{6}B}  \right)^4}=0 \label{C8}.
\end{equation}

The parameters $B$ and $A$ can be obtained as follows.

\begin{equation}
B=-\frac{\sqrt[4]{2}}{\sqrt{3}} \label{C9}.
\end{equation}

\begin{equation}
A=2^{\frac{3}{4}} \label{C10}.
\end{equation}

By substituting (\ref{C9}) into (\ref{C5}), the parameter $\phi$ is obtained from the following equation.

\begin{equation}
\mathrm{sleafh}_2 \left( -\frac{\sqrt[4]{2}}{\sqrt{3}} +\phi \right)
=\sqrt[4]{2} \label{C11}.
\end{equation}

\begin{equation}
\phi=a\mathrm{sleafh}_2 \left( \sqrt[4]{2}   \right)
+\frac{\sqrt[4]{2}}{\sqrt{3}} \label{C12}.
\end{equation}

The prefix $a$ of the hyperbolic leaf function $\mathrm{sleafh}_2(t)$ represents the inverse function. 

\appendix
\def\thesection{Appendix}
\section{D}
\label{D}
The equation for obtaining the period is now derived. By multiplying Eq. (\ref{3.2.2}) with $\mathrm{d}\theta/\mathrm{d}t$, the following equation is obtained.

\begin{equation}
\frac{\mathrm{d}^2 \theta}{\mathrm{d}t^2} \frac{\mathrm{d} \theta}{\mathrm{d}t}+\frac{g}{L} \theta \frac{\mathrm{d} \theta}{\mathrm{d}t}=0 \label{D1}.
\end{equation}

The initial conditions are the Eq. (\ref{3.1.4}) and the Eq. (\ref{3.1.5}). By integrating the above equation, the following equation is obtained.

\begin{equation}
\left( \frac{\mathrm{d} \theta}{\mathrm{d}t} \right)^2 
=\omega^2 \left(4 - \theta^2 \right) \label{D2}.
\end{equation}

where the parameter $\omega$ is given by Eq. (\ref{3.1.3}). When the variables are separated to isolate $\theta$ and $t$, the following equation is obtained.

\begin{equation}
\omega \mathrm{d}t =\pm \frac{\mathrm{d} \theta}{\sqrt{4-\theta^2 }} \label{D3}.
\end{equation}

Where the initial condition is $\theta=2$ at $t = 0$. As time progress, the mass of the pendulum moves to the position $\theta=0$. The time required for this movement represents a quarter period, where one period is defined as $T$. By integrating the above equation, the following equation is obtained. Since the period $T$ is positive, a negative sign $-$ is applied to Eq. (\ref{D3})

\begin{equation}
\int_{0}^{\frac{T}{4}} \omega \mathrm{d}t =-\int_{2}^{0} \frac{\mathrm{d} \theta}{\sqrt{4-\theta^2 }}  \label{D4}.
\end{equation}

The integral equation for computing the period is obtained.

\begin{equation}
T =\frac{4}{\omega} \int_{0}^{2} \frac{\mathrm{d} \theta}{\sqrt{4-\theta^2 }}  \label{D5}.
\end{equation}

Similarly, the period from Eq. (\ref{3.2.5}) is obtained using the following equation.

\begin{equation}
T =\frac{4}{\omega} \int_{0}^{2} 
\frac{\mathrm{d} \theta}{\sqrt{\frac{8}{3}-\theta^2+\frac{\theta^4}{12} }}  \label{D6}.
\end{equation}

Similarly, the period from Eq. (\ref{3.2.1}) is obtained using the following equation.

\begin{equation}
T =\frac{2 \sqrt{2} }{\omega} \int_{0}^{2} \frac{\mathrm{d} \theta}{\sqrt{\mathrm{cos}(\theta)-\mathrm{cos}(2)}}  \label{D7}.
\end{equation}

\appendix
\def\thesection{Appendix}
\section{E}
\label{E}
The constant parameter $\pi_2$ is obtained from the following equation \cite{Kaz_sl} \cite{Kaz_cl}.

\begin{equation}
\pi_2 =2 \int_{0}^{1} \frac{ 1 }{ \sqrt{ 1-t^4 } }  \label{E1}.
\end{equation}

\appendix
\def\thesection{Appendix}
\section{F}
\label{F}

The following equation can be obtained from Eq. (\ref{A4}) by setting $B=1$, $\phi=0$, $\omega=1$ and $t=\pi_2/2$.

\begin{equation}
\mathrm{cos} \left( 2 \int_0^{\frac{\pi_2}{2}} \mathrm{cleaf}_2(u) \mathrm{d}u \right)=\left(\mathrm{cleaf}_2 \left( \frac{\pi_2}{2} \right) \right)^2=0
\label{F1}.
\end{equation}

The range of possible phases of Eq. (\ref{F1}) is as follows \cite{Kaz_duf}.

\begin{equation}
-\frac{\pi}{4} \leqq \int_0^{\frac{\pi_2}{2}} \mathrm{cleaf}_2(u) \mathrm{d}u  \leqq \frac{\pi}{4}
\label{F2}.
\end{equation}

Therefore, the following equation is obtained.

\begin{equation}
2 \int_0^{\frac{\pi_2}{2}} \mathrm{cleaf}_2(u) \mathrm{d}u  = \frac{\pi}{2}
\label{F3}.
\end{equation}

Here, the parameter $u$ is replaced by the equation $u=v/\sqrt{3}$. The equation $u = 0$ is replaced by $v = 0$. The equation $u = \pi_2/2$ is replaced by $v = \sqrt{3} \pi_2/2$. The following equation is then obtained.

\begin{equation}
2 \int_0^{\frac{\sqrt{3}}{2} \pi_2 } 
\mathrm{cleaf}_2 \left( \frac{1}{\sqrt{3}}v \right)\frac{1}{\sqrt{3}} \mathrm{d}v
= \frac{\pi}{2}
\label{F4}.
\end{equation}

Using the above equation, the conversion in Eq. (\ref{B3}) is obtained.

\appendix
\def\thesection{Appendix}
\section{G}
\label{G}

The leaf functions and hyperbolic leaf functions based on the basis $n=1$ are as follows:
\begin{equation}
\mathrm{sleaf}_{1}(t)=\mathrm{sin}(t).
\end{equation}
\begin{equation}
\mathrm{cleaf}_{1}(t)=\mathrm{cos}(t).
\end{equation}
\begin{equation}
\mathrm{sleafh}_{1}(t)=\mathrm{sinh}(t).
\end{equation}
\begin{equation}
\mathrm{cleafh}_{1}(t)=\mathrm{cosh}(t).
\end{equation}
The lemniscate functions are presented by Gauss \cite{Gauss, roy}. The relation equations between this function and the leaf function are as follows:
\begin{equation}
\mathrm{sleaf}_{2}(t)=\mathrm{sl}(t).
\end{equation}
\begin{equation}
\mathrm{cleaf}_{2}(t)=\mathrm{cl}(t).
\end{equation}
\begin{equation}
\mathrm{sleafh}_{2}(t)=\mathrm{slh}(t) \label{slh}.
\end{equation}
The definition of the function $\mathrm{slh}(t)$ of Eq.(\ref{slh}) can be confirmed by references \cite{Car2, Neu}. For the function corresponding to the hyperbolic leaf function $\mathrm{cleafh}_2(t)$, no clear description can be found in the literature.

\nocite{*}

\bibliographystyle{unsrt}  
\bibliography{references}  

\end{document}